\documentclass[12pt]{article}

\RequirePackage{amsmath,amsthm,amsfonts,amssymb}
\RequirePackage[colorlinks,citecolor=blue,urlcolor=blue]{hyperref}
\RequirePackage{graphicx}

\usepackage{float}
\usepackage{enumitem}
\usepackage{url}
\usepackage{bbm}
\usepackage{natbib}
\usepackage[nottoc,numbib]{tocbibind}
\usepackage{booktabs,caption,subcaption}
\usepackage[flushleft]{threeparttable}
\usepackage{multirow}
\newcommand{\ra}[1]{\renewcommand{\arraystretch}{#1}}
\usepackage{xr}
\usepackage{color}
\usepackage{tikz}
\usetikzlibrary{positioning}
\usetikzlibrary{trees}
\usepackage{microtype}

\usepackage[margin=1in]{geometry}

\newcommand{\bA}{\boldsymbol{A}}
\newcommand{\mA}{\mathcal{A}}
\newcommand{\ba}{\boldsymbol{a}}

\newcommand{\bC}{\boldsymbol{C}}

\newcommand{\E}{\mathbb{E}}

\newcommand{\bI}{\boldsymbol{I}}
\newcommand{\bJ}{\boldsymbol{J}}
\newcommand{\bj}{\boldsymbol{j}}

\newcommand{\mN}{\mathcal{N}}

\newcommand{\bone}{\boldsymbol{1}}

\newcommand{\bS}{\boldsymbol{S}}
\newcommand{\mS}{\mathcal{S}}
\newcommand{\bs}{\boldsymbol{s}}
\newcommand{\bT}{\boldsymbol{T}}

\newcommand{\V}{\mathbb{V}}
\newcommand{\bV}{\boldsymbol{V}}
\newcommand{\bv}{\boldsymbol{v}}
\newcommand{\bW}{\boldsymbol{W}}

\newcommand{\bX}{\boldsymbol{X}}

\newcommand{\bzero}{\boldsymbol{0}}

\newcommand{\bbeta}{\boldsymbol{\beta}}

\newcommand{\bgamma}{\boldsymbol{\gamma}}

\newcommand{\bphi}{\boldsymbol{\phi}}
\newcommand{\bPsi}{\boldsymbol{\Psi}}
\newcommand{\bpsi}{\boldsymbol{\psi}}
\newcommand{\brho}{\boldsymbol{\rho}}

\newcommand{\btheta}{\boldsymbol{\theta}}

\newtheorem{theorem}{Theorem}
\newtheorem{corollary}{Corollary}

\allowdisplaybreaks

\begin{document}

\title{\bf Distributed model building and recursive integration for big spatial data modeling}
\author{Emily C. Hector\thanks{Hector and Reich were supported by a grant from the National Science Foundation (DMS2152887). Hector was also supported by a Faculty Research and Professional Development Award from North Carolina State University.}\hspace{.2cm}\\
Department of Statistics, North Carolina State University\\
Brian J. Reich\\Department of Statistics, North Carolina State University\\
Ani Eloyan\\Department of Biostatistics, Brown University}
\date{}
\maketitle
  
\bigskip
\begin{abstract}
\noindent Motivated by the need for computationally tractable spatial methods in neuroimaging studies, we develop a distributed and integrated framework for estimation and inference of Gaussian process model parameters with ultra-high-dimensional likelihoods. We propose a shift in viewpoint from whole to local data perspectives that is rooted in distributed model building and integrated estimation and inference. The framework’s backbone is a computationally and statistically efficient integration procedure that simultaneously incorporates dependence within and between spatial resolutions in a recursively partitioned spatial domain. Statistical and computational properties of our distributed approach are investigated theoretically and in simulations. The proposed approach is used to extract new insights on autism spectrum disorder from the Autism Brain Imaging Data Exchange.
\end{abstract}

\noindent%

\section{Introduction}
\label{sec:introduction}

The proposed methods are motivated by the investigation of differences in brain functional organization between people with Autism Spectrum Disorder (ASD) and their typically developing peers. The Autism Brain Imaging Data Exchange (ABIDE) neuroimaging study of resting-state functional Magnetic Resonance Imaging (rfMRI) aggregated and publicly shared neuroimaging data on participants with ASD and neurotypical controls from 16 international imaging sites. rfMRI measures blood oxygenation in the absence of a stimulus or task and characterizes intrinsic brain activity \citep{Fox-Raichle}. Relationships between activated brain regions of interest (ROIs) during rest can be characterized by functional connectivity between ROIs using rfMRI data. Functional connectivity between two ROIs is typically estimated from the subject-specific correlation constructed from the rfMRI time series \citep{He-etal}. To avoid the computational burden of storing and modeling a large connectivity matrix, the rfMRI time series is averaged across voxels in each ROI before computing the cross-ROI correlation and modeling its relationship with covariates. Other approaches for quantifying covariate effects on correlation are primarily univariate \citep[see, e.g.,][]{Shehzad-etal}. These standard approaches are substantially underpowered to detect small effect sizes and obscure important variation within ROIs. While ABIDE data have led to some evidence that ASD can be broadly characterized as a brain dysconnection syndrome, findings have varied across studies \citep{Uddin-etal, Alaerts-etal, DiMartino-etal-2014}. Computationally and statistically efficient estimation of functional connectivity maps is essential in identifying ASD dysconnections. In this paper, we show how to estimate the effect of covariates including ASD on over 6.5 million within- and between-ROI correlations in approximately $5$ hours.

The first component of our efficient approach is to leverage the spatial dependence within- and between all voxels in ROIs using Gaussian process models \citep{Cressie, Banerjee-etal-2014, Cressie-Wikle}. Denote the rfMRI outcomes $\{y_i(\bs)\}_{i=1}^N$ for $N$ independent samples at one of $S$ voxels $\bs \in \mS$. The joint distribution of $y_i(\mS)$ is assumed multivariate Gaussian and known up to a vector of parameters of interest $\btheta$. Without further modeling assumptions or dimension reduction techniques, maximum likelihood estimation has memory and computational complexity $O(NS^2)$ and $O(NS^3)$ respectively due to the $S$-dimensional covariance matrix. For inference on $\btheta$ when $S$ is large, the crux of the problem is to adequately model the spatial dependence without storing or inverting a large covariance matrix. 

This problem has received considerable attention \citep{Sun-etal, Bradley-etal, Heaton-etal-2019, Liu-etal}. Solutions include, for example, the composite likelihood (CL) \citep{Lindsay, Varin-Reid-Firth}  the nearest-neighbour Gaussian process \citep{Datta-etal-a, Finley-etal-2019}, spectral methods \citep{Fuentes}, tapered covariance functions \citep{Furrer-etal-2006, Kaufman-etal, Stein-2013}, low-rank approximations \citep{Zimmerman, Cressie-Johannesson, Nychka-etal}, and combinations thereof \citep{Sang-Huang}. \cite{Stein-2013} details shortcomings of the aforementioned methods, which are primarily related to loss of statistical efficiency due to simplification of the covariance matrix or its inverse. Moreover, these methods remain computationally burdensome when $S$ is very large, a problem which is further exacerbated when the covariance structure is nonstationary due to covariate effects on the spatial correlation parameters. 

We propose a new distributed and integrated framework for big spatial data analysis. We consider a recursive partition of the spatial domain $\mS$ into $M$ nested resolutions with disjoint sets of spatial observations at each resolution, and build local, fully specified distributed models in each set at the highest spatial resolution. The main technical difficulty arises from integrating inference from these distributed models over two levels of dependence: between sets in each resolution, and between resolutions. Simultaneously incorporating dependence between all sets and resolutions results in a high-dimensional dependence matrix that is computationally prohibitive to handle. The main contribution of this paper is a recursive estimator that integrates inference over all sets and resolutions by alternating between levels of dependence at each integration step. We also propose a sequential integrated estimator that is asymptotically equivalent to the recursive integrated estimator but reduces the computational burden of recursively integrating over multiple resolutions. The resulting Multi-Resolution Recursive Integration (MRRI) framework is flexible, statistically efficient, and computationally scalable through its formulation in the MapReduce paradigm.

The rest of this paper is organized as follows. Section \ref{sec:method} establishes the formal problem setup, and describes the distributed model building step and the recursive integration scheme for the proposed MRRI framework. Section \ref{sec:simulations} evaluates the proposed frameworks with simulations. Section \ref{sec:application} presents the analysis of data from ABIDE. Proofs, additional results, ABIDE information and an R package are provided in the supplement.

\section{Recursive Model Integration Framework}
\label{sec:method}

\subsection{Problem set-up}
\label{subsec:setup}

Suppose we observe $y_i(\bs)=\alpha_i(\bs) + \epsilon_i(\bs)$ the $i$th observation at location $\bs \in \mathbb{R}^d$, $i=1, \ldots, N$ independently and $\bs \in \mS$ a set of $S$ locations, where $\alpha_i(\cdot)$ characterizes the spatial variations and $\epsilon_i(\cdot)$ is an independent normally distributed measurement error with mean $0$ and variance $\sigma^2$ independent of $\alpha_i(\cdot)$. Further, suppose for each independent replicate $i$ that we observe $q$ explanatory variables $\bX_i(\bs) \in \mathbb{R}^q$. We denote by $\mS=\{\bs_j\}_{j=1}^S$ the set of locations and define $y_i(\mS)=\{y_i(\bs_j)\}_{j=1}^S$ and $\bX_i(\mS)=\{\bX_i(\bs_j)\}_{j=1}^S$. 

\begin{figure}[h]
\centering
\includegraphics[width=0.25\textwidth]{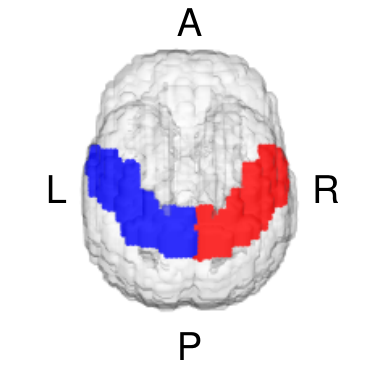}
\caption{Left (L, blue) and right (R, red) precentral gyri. A: Anterior; P: Posterior.\label{f:lr_precent_gyri}}
\end{figure}

In Section \ref{sec:application}, $\mS$ is the set of voxels in the left and right precentral gyri, visualized in Figure \ref{f:lr_precent_gyri}, and $d=3$. Outcomes $\{y_i(\bs)\}_{i=1}^N$ consist of the thinned rfMRI time series at voxel $\bs$ for ABIDE participants passing quality control, where the thinning removes every second time point to remove autocorrelation; autocorrelation plots are provided in the supplement. Thinning avoids the bias occasionally introduced by other approaches \citep{Monti} and respects usual rfMRI autocorrelation assumptions \citep{Arbabshirani-etal}. Specifically, for participant $n\in \{1, \ldots, 774\}$ and voxel $\bs \in \mS$ with a (centered and standardized) time series of length $2T_n$, the outcome $y_i(\bs)$ consists of the rfMRI outcome at time point $t \in \{ 2,4, \ldots, 2T_n\}$, where $i= t + \sum_{n^\prime < n} T_{n^\prime}$ indexes the participants and time points. This thinning results in $N=75888$ observations of $y_i(\bs)$ that are independent across $i=1, \ldots, N$. While we focus on two ROIs for simplicity of exposition, our approach generalizes to multiple ROIs. Variables $\bX_i(\bs)=\bX_i$ consist of an intercept, ASD status, age, sex and the age by ASD status interaction.

We assume that $\alpha_i(\cdot)$ is a Gaussian Process with mean function $\mu\{ \cdot; \bX_i(\cdot), \bbeta \}$ and positive-definite covariance function $C_{\alpha}\{ \cdot,\cdot; \bX_i(\cdot), \bgamma\}$. We further assume that $\mu\{ \mS; \bX_i(\mS), \bbeta \}$ is known up to a $q_1$-dimensional vector of parameters $\bbeta$, and $C_{\alpha}\{ \mS, \mS; \bX_i(\mS), \bgamma \}$ is known up to a $q_2$-dimensional vector of parameters, $\bgamma$: $C_{\alpha}\{ \bs_1, \bs_2; \allowbreak \bX_i(\bs_1), \bX_i(\bs_2), \bgamma\}=\mbox{Cov}\{ \alpha_i(\bs_1), \alpha_i(\bs_2) \}$. We allow the covariance to depend on $\bX_i(\mS)$, resulting in a nonstationary covariance structure. We define $\btheta=(\bbeta, \bgamma, \sigma^2) \in \mathbb{R}^p$ the parameter of interest, $p=q_1+q_2+1$. In Section \ref{sec:application}, $C_{\alpha}$ borrows from \cite{Paciorek-Schervish} and takes the form
\begin{align}
2^{\frac{d}{2}} \tau(\bs_j, \bs_{j^\prime}) \Biggl\{ \frac{\rho_{ij^\prime} \rho_{ij}}{(\rho_{ij^\prime} + \rho_{ij})^2} \Biggr\}^{\frac{d}{4}}
\exp \Biggl[ \frac{- 2(\bs_{j^\prime} - \bs_j)^\top (\bs_{j^\prime} - \bs_j)}{\rho_{ij^\prime} + \rho_{ij}} \Biggr],
\label{e:ABIDE-cov}
\end{align}
where $\rho_{ij}=\exp\{ \bX_i^\top \brho(\bs_j)\}$ models the effect of $\bX_i$ on the correlation between voxels through $\brho(\bs_j) \in \mathbb{R}^q$. Both $\brho(\bs_j)$ and $\tau(\bs_j, \bs_{j^\prime})$ are specified in Section \ref{sec:application}. 

The distribution of $y_i(\mS)$ is $S$-variate Gaussian with mean $\mu\{ \mS; \bX_i(\mS), \bbeta \}$ and covariance $\bC\{ \mS, \mS; \bX_i(\mS), \bgamma, \sigma^2\}=[ C_{\alpha}\{\bs_r, \bs_t; \bX_i(\bs_r), \bX_i(\bs_t), \bgamma\} ]_{r,t=1}^{S,S} + \sigma^2 \bI_S$. To overcome the difficulty of an intractable likelihood when $S$ is large, we borrow ideas from the CL, multi-resolution approximations (MRA) and generalized method of moments (GMM).

\subsection{Recursive Integration Framework} 
\label{subsec:distributed}

\subsubsection{A Shift from Whole to Local Data Perspectives}

The CL \citep{Lindsay, Varin-Reid-Firth} divides $\mS$ into $K$ (potentially overlapping) subsets, builds well-specified local models on the $K$ subsets, and integrates them using working independence assumptions. The CL is attractive because it balances statistical and computational efficiencies, and the maximum CL estimator is consistent and asymptotically normal under mild regularity conditions \citep{Cox-Reid}. The main difficulty in its construction is the choice of $K$, which regulates both the number and the dimensionality of the marginal densities. Generally, large $K$ is preferred as it alleviates the modeling difficulties and computational burden associated with specifying and evaluating multivariate densities. Large $K$, however, can result in the evaluation of a large number of low-dimensional marginals, which is computationally expensive and inefficient. When $S$ is truly large, no choice of $K$ is adequately statistically or computationally efficient, since the number of margins is large and their dimension remains high. To achieve a small number of low-dimensional sets, we propose a recursive partition of $\mS$ into multiple resolutions, with multiple sets at each resolution.

This idea is, at first blush, similar to MRA models \citep{Nychka-etal, Katzfuss, Katzfuss-Gong}. Where they integrate from global to local levels, however, we build local models at the highest resolution and recursively integrate inference from the highest to the lowest resolution. Unlike MRA models, we incorporate dependence within all spatial resolutions using the GMM \citep{Hansen} framework. 

The GMM minimizes a weighted quadratic form of estimating functions and provides an intuitive mechanism for incorporating dependence between local models. Recently, \cite{Hector-Song-JASA, Manschot-Hector, Hector-Reich} proposed closed-form meta-estimators for integrating estimators from dependent analyses that are asymptotically as statistically efficient as the most efficient GMM estimators but that avoid computationally expensive iterative minimization of the GMM objective function. Extending this framework to recursively partitioned spatially dependent observations, however, leads to inversion of a dependence matrix for all sets of observations that is unfortunately high-dimensional, negating the gain in computation afforded by the partition. In what follows, we propose a new weighting scheme to optimally weight GMM estimating functions and reduce their dimension for computationally tractable recursive integration of dependent models.

\subsubsection{Partitioning the Spatial Domain} 

\begin{figure}[h!]
\centering
\vspace{1em}
\begin{tikzpicture}
\node (pic) at (0,0) {\includegraphics[width=\textwidth]{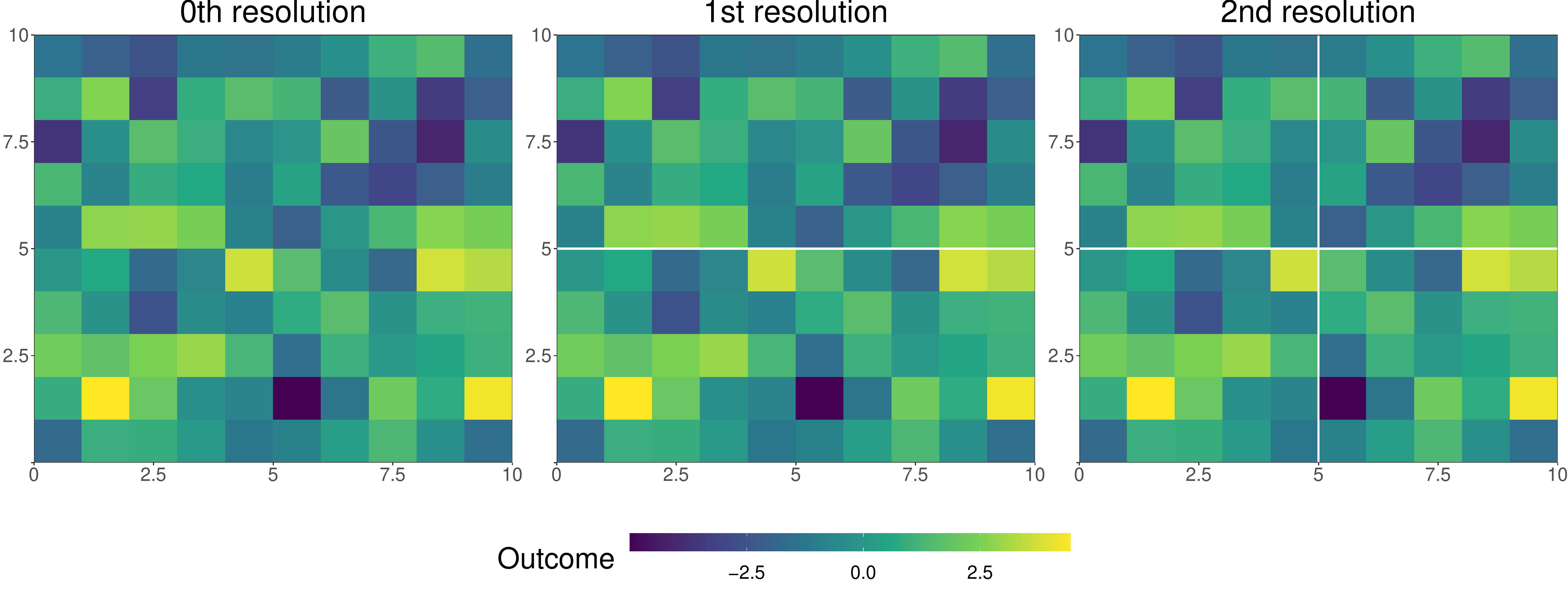}};
\node (A0) at (-5.2,0.6) {\color{white}$\mA_0$};
\node (A1) at (0.1,1.7) {\color{white}$\mA_1$};
\node (A2) at (0.1,-0.7) {\color{white}$\mA_2$};
\node (A11) at (4.3,1.7) {\color{white}$\mA_{11}$};
\node (A12) at (6.8,1.7) {\color{white}$\mA_{12}$};
\node (A21) at (4.3,-0.7) {\color{white}$\mA_{21}$};
\node (A22) at (6.8,-0.7) {\color{white}$\mA_{22}$};
\end{tikzpicture}
\caption{Example partition of observations on a two-dimensional spatial domain, $M=2$.\label{toy-example}}
\end{figure}
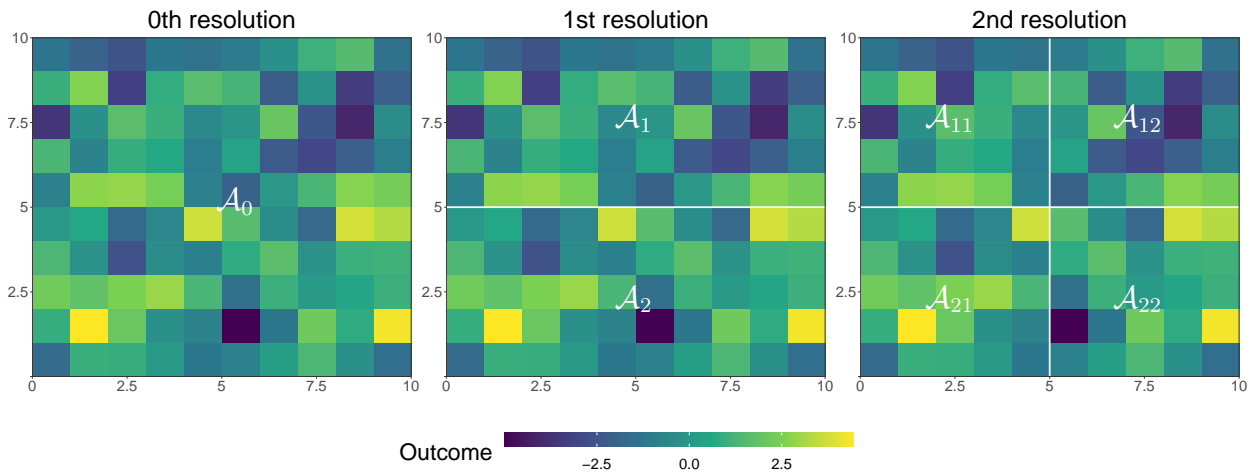

Let $\otimes 2$ denote the outer product of a vector with itself, namely $\ba^{\otimes 2}=\ba \ba^\top$. We adopt the notation of \cite{Katzfuss} to describe the recursive partitioning of $\mS$. Denote $\mA_0=\mS$ and partition $\mS$ into $K_1$ (disjoint) regions $\{ \mA_1, \ldots, \mA_{K_1}\}$ that are again partitioned into $K_2$ (disjoint) subregions $\{ \mA_{k_11}, \ldots, \mA_{k_1K_2}\}_{k_1=1}^{K_1}$, and so on up to level $M$, i.e. $\mA_{k_1 \ldots k_{m-1}}=\cup_{k_m=1, \ldots, K_m} \mA_{k_1 \ldots k_{m-1} k_m}$, $k_{m-1} \in \{1, \ldots, K_{m-1}\}$, $m=0, \ldots, M$, where $K_0=1$. For completeness, let $k_0=0$. Figure \ref{toy-example} illustrates an example of a recursive partition of observations on a two-dimensional spatial domain for $M=2$ resolutions. For resolution $m \in \{1, \ldots, M\}$, denote $S_{k_1\ldots k_m}$ the size of $\mA_{k_1\ldots k_m}$, with $S_0=S$. Further define $k_{\max}=\max k_m$ with the maximum taken over $k_m=1, \ldots, K_m$ and $m=1, \ldots, M$. Values of $M$ and $K_m$ should be chosen so that $S_{k_1\ldots k_M}$ and $K_m, m=1, \ldots, M$ are relatively small compared to $S_0$, and we generally recommend $S_{k_1\ldots k_M} \geq 25$. The literature is replete with methods for choosing partitions; see \cite{Heaton-etal-2019} for an excellent review. In Section \ref{sec:application}, we partition voxels $\mS$ based on nearest neighbours with $M=3$ resolutions.

\subsubsection{Local Model Specification} 

At resolution $M$ for $k_M \in \{1, \ldots, K_M \}$, the likelihood of the data in set $\mA_{k_1 \ldots k_M}$ is given by $y_i(\mA_{k_1\ldots k_M}) \sim \mN [ \mu \{ \mA_{k_1\ldots k_M}; \bX_i(\mA_{k_1\ldots k_M}), \bbeta\}; \bC\{ \mA_{k_1\ldots k_M}, \mA_{k_1\ldots k_M}; \bX_i(\mA_{k_1\ldots k_M}), \bgamma, \sigma^2\} ]$. We model the mean of the spatial process through $\mu\{ \mA_{k_1\ldots k_M}; \bX_i(\mA_{k_1\ldots k_M}), \allowbreak \bbeta\}=\bX^\top_i(\mA_{k_1\ldots k_M}) \bbeta$. In Section \ref{sec:application}, $\mu\{ \bs_j; \bX_i(\bs_j), \bbeta\}=\beta$ for $\bs_j \in \mA_{k_1\ldots k_M}$ due the centering of each time series, and $\bC$ is the nonstationary covariance function in \eqref{e:ABIDE-cov} that models the effect of covariates on functional connectivity within and between the left and right precentral gyri. Due to the recursive partitioning, $S_{k_1\ldots k_M}$ is small and the full likelihood tractable. Denote the log-likelihood of the data in $\mA_{k_1\ldots k_M}$ as $\ell_{k_1\ldots k_M}(\btheta)$ and the score function as $\bPsi_{k_1\ldots k_M}(\btheta)=\nabla_{\btheta} \ell_{k_1\ldots k_M}(\btheta)=\sum_{i=1}^N \bpsi_{i,k_1\ldots k_M}(\btheta)$. The maximum likelihood estimator (MLE) $\widehat{\btheta}_{k_1\ldots k_M}$ of $\btheta$ in $\mA_{k_1\ldots k_M}$ is obtained by maximizing the log-likelihood, or equivalently solving $\bPsi_{k_1\ldots k_M}(\btheta)=\boldsymbol{0}$.

\subsection{Recursive Model Integration}
\label{subsec:recursive}

\subsubsection{Generalized Method of Moments Approach} 
\label{subsubsec:GMM}

We now wish to recursively integrate the local models at each resolution. For $m=M, \ldots, 0$, let $k_m \in \{1, \ldots, K_m\}$ denote the index for the sets at resolution $m$. We describe a recursive GMM approach that re-estimates $\btheta$ at each resolution $M-1$ based on an optimally weighted form of the GMM equation from the higher resolution $M$. At resolution $M-1$, we define $K_{M-1}$ estimating functions based on the score functions from resolution $M$:
\begin{equation*}
\begin{split}
\bpsi_{i,k_1\ldots k_{M-1}}(\btheta)=\{ \bpsi_{i,k_1\ldots k_{M-1} k_M}(\btheta) \}_{k_M=1}^{K_M} \in \mathbb{R}^{pK_M}, \quad k_{M-1}=1, \ldots, K_{M-1},\\
\bPsi_{k_1\ldots k_{M-1}}(\btheta)=\{ \bPsi_{k_1\ldots k_{M-1} k_M}(\btheta) \}_{k_M=1}^{K_M} \in \mathbb{R}^{pK_M}, \quad k_{M-1}=1, \ldots, K_{M-1}.
\end{split}
\end{equation*}
The estimating function $\bPsi_{k_1\ldots k_{M-1}}(\btheta)$ over-identifies $\btheta \in \mathbb{R}^p$, i.e. there are several estimating functions for each parameter. Following \cite{Hansen}'s GMM, we propose to minimize the quadratic form $Q_{k_1\ldots k_{M-1}}(\btheta)=\bPsi^\top_{k_1\ldots k_{M-1}}(\btheta) \bW \bPsi_{k_1\ldots k_{M-1}}(\btheta)$, with a positive semi-definite weight matrix $\bW$. \cite{Hansen} showed that the optimal choice of $\bW$ is the inverse of the covariance  of $\bPsi_{k_1\ldots k_{M-1}}(\btheta)$, which can be consistently estimated by $N^{-1}\bV_{k_1\ldots k_{M-1}}(\widehat{\btheta}_{k_1\ldots k_M})$,
\begin{align*}
\bV_{k_1\ldots k_{M-1}}(\widehat{\btheta}_{k_1\ldots k_M})= \sum \limits_{i=1}^N \bigl\{ \bpsi_{i,k_1\ldots k_{M-1}}(\widehat{\btheta}_{k_1\ldots k_M}) \bigr\}^{\otimes 2}
\end{align*}
the variability matrix, and $\bpsi_{i,k_1\ldots k_{M-1}}(\widehat{\btheta}_{k_1\ldots k_M})=\{ \bpsi_{i,k_1\ldots k_M}(\widehat{\btheta}_{k_1\ldots k_{M-1} k_M}) \}_{k_M=1}^{K_M} $. The matrix $N^{-1}\bV_{k_1\ldots k_{M-1}}(\widehat{\btheta}_{k_1\ldots k_M})$ estimates dependence between score functions from all sets in $\mA_{k_1\ldots k_{M-1}}$ and thus captures dependence between sets $\{ \bA_{k_1\ldots k_M} \}_{k_M=1}^{K_M}$. This choice is ``Hansen'' optimal in the sense that, for all possible choices of $\bW$,
\begin{align}
\widehat{\btheta}_{GMM}=\arg \min_{\btheta} Q^*_{k_1\ldots k_{M-1}}(\btheta)=\arg \min \bPsi^\top_{k_1\ldots k_{M-1}}(\btheta) \bV^{-1}_{k_1\ldots k_{M-1}}(\widehat{\btheta}_{k_1\ldots k_M}) \allowbreak \bPsi_{k_1\ldots k_{M-1}}(\btheta)
\label{e:opt-GMM}
\end{align}
has minimum variance among estimators $\arg \min_{\btheta} Q_{k_1\ldots k_{M-1}}(\btheta)$. The computation of $Q^*_{k_1\ldots k_{M-1}}(\btheta)$ in \eqref{e:opt-GMM} requires inversion of the $(pK_M)\times (pK_M)$ dimensional matrix $\bV^{-1}_{k_1\ldots k_{M-1}}(\widehat{\btheta}_{k_1\ldots k_M})$, substantially smaller than inverting the $(pS_{k_1\ldots k_{M-1}})\times(pS_{k_1\ldots k_{M-1}})$ dimensional matrix in the direct evaluation of the likelihood on set $\mA_{k_1\ldots k_{M-1}}$. This yields a faster computation than a full likelihood approach on $\mA_{k_1\ldots k_{M-1}}$. The iterative minimization in \eqref{e:opt-GMM}, however, can be time consuming because it requires computation of the score functions $\bPsi_{k_1\ldots k_M}$ at each iteration. 

In the spirit of \cite{Hector-Song-JASA}, a closed-form meta-estimator asymptotically equivalent to $\widehat{\btheta}_{GMM}$ in \eqref{e:opt-GMM} that is more computationally attractive is given by 
\begin{align}
\widehat{\btheta}_{k_1\ldots k_{M-1}}&=\bJ^{-1}_{k_1\ldots k_{M-1}}(\widehat{\btheta}_{k_1\ldots k_M}) \bS_{k_1\ldots k_{M-1}}(\widehat{\btheta}_{k_1\ldots k_M}) \bV^{-1}_{k_1\ldots k_{M-1}}(\widehat{\btheta}_{k_1\ldots k_M}) \bT_{k_1\ldots k_{M-1}}(\widehat{\btheta}_{k_1\ldots k_M}),
\label{e:meta-m-1}
\end{align}
where $\bS_{k_1\ldots k_{M-1}}(\widehat{\btheta}_{k_1\ldots k_M})=-\{ \nabla_{\btheta} \bPsi_{k_1\ldots k_{M-1}}(\btheta) \rvert_{\btheta=\widehat{\btheta}_{k_1\ldots k_M}} \}^\top \in \mathbb{R}^{p\times pK_M}$,
\begin{align*}
\bT_{k_1\ldots k_{M-1}}(\widehat{\btheta}_{k_1\ldots k_M})&=\left\{
\bS^\top_{k_1\ldots k_{M-1} k^\prime_M}(\widehat{\btheta}_{k_1\ldots k_M}) \widehat{\btheta}_{k_1\ldots k_{M-1} k^\prime_M} 
 \right\}_{k^\prime_M=1}^{K_M} \in \mathbb{R}^{pK_M},\\
\bJ_{k_1\ldots k_{M-1}}(\widehat{\btheta}_{k_1\ldots k_M})&=\bS_{k_1\ldots k_{M-1}}(\widehat{\btheta}_{k_1\ldots k_M}) \bV^{-1}_{k_1\ldots k_{M-1}}(\widehat{\btheta}_{k_1\ldots k_M}) \bS^\top_{k_1\ldots k_{M-1}}(\widehat{\btheta}_{k_1\ldots k_M}) \in \mathbb{R}^{p\times p}.
\end{align*}
The sensitivity matrix $\bS_{k_1\ldots k_{M-1}}(\widehat{\btheta}_{k_1\ldots k_M})$ can be efficiently computed by Bartlett's identity.

When updating to the next resolution $M-2$, it is tempting to proceed again through the GMM approach and to stack estimating functions $\{ \bPsi_{k_1 \ldots k_{M-1}}\}_{k_{M-1}=1}^{K_{M-1}}$. This would, however, result in a vector of over-identified estimating functions on $\btheta$ of dimension $pK_{M-1}K_M$, and therefore inversion of a $(pK_{M-1}K_M)\times(pK_{M-1}K_M) $ dimensional covariance matrix. Recursively proceeding this way would lead to the computationally prohibitive inversion of a $(pK_1\ldots K_M)\times (pK_1\ldots K_M)$ dimensional matrix at resolution $m=0$. To avoid this difficulty, we propose weights in the spirit of optimal estimating function theory \citep{Heyde}.

\subsubsection{Weighted Over-Identified Estimating Functions} 
\label{subsubsec:weighted}

We define new estimating functions at resolution $m=M-2, \ldots, 0$ that optimally weight the estimating functions from resolution $m+1$,
\begin{equation}
\begin{split}
\widetilde{\bpsi}_{i,k_1\ldots k_{M-1}}(\btheta)&= - \bS_{k_1\ldots k_{M-1}}(\widehat{\btheta}_{k_1\ldots k_{M-1}}) \bV^{-1}_{k_1\ldots k_{M-1}}(\widehat{\btheta}_{k_1\ldots k_{M-1}}) \bpsi_{i,k_1\ldots k_{M-1}}(\btheta) \in \mathbb{R}^p\\
\widetilde{\bPsi}_{k_1\ldots k_{M-1}}(\btheta)&= - \bS_{k_1\ldots k_{M-1}}(\widehat{\btheta}_{k_1\ldots k_{M-1}}) \bV^{-1}_{k_1\ldots k_{M-1}}(\widehat{\btheta}_{k_1\ldots k_{M-1}}) \bPsi_{k_1\ldots k_{M-1}}(\btheta) \in \mathbb{R}^p,
\label{e:weighted-EE}
\end{split}
\end{equation}
where $\bS^\top_{k_1\ldots k_{M-1}}(\widehat{\btheta}_{k_1\ldots k_{M-1}})$ and $\bV_{k_1\ldots k_{M-1}}(\widehat{\btheta}_{k_1\ldots k_{M-1}})$
are recomputed so as to evaluate the sensitivity and variability at the estimator from resolution $M-1$. This formulation can also be viewed as the optimal projection of the estimating function $\bPsi_{k_1\ldots k_{M-1}}(\btheta)$ onto the parameter space of $\btheta$ \citep{Heyde}. Stacking $\{\widetilde{\bPsi}_{k_1\ldots k_{M-1}}(\btheta)\}_{k_{M-1}=1}^{K_{M-1}}$ yields $\bPsi_{k_1\ldots k_{M-2}}(\btheta)$, a $pK_{M-1}$ dimensional vector of over-identifying estimating functions on $\btheta$. Defining $\bV_{k_1\ldots k_{M-2}}(\widehat{\btheta}_{k_1\ldots k_{M-1}})$ the sample covariance of $\bPsi_{k_1\ldots k_{M-2}}(\btheta)$ evaluated at $\widehat{\btheta}_{k_1\ldots k_{M-1}}$, one can again define the closed-form meta-estimator $\widehat{\btheta}_{k_1\ldots k_{M-2}}$ in the fashion of \eqref{e:meta-m-1}. This requires inversion of a $(pK_{M-1})\times(pK_{M-1})$-dimensional matrix, substantially smaller than inverting the $S_{k_1\ldots k_{M-2}}\times S_{k_1\ldots k_{M-2}}$-dimensional matrix in the full likelihood on set $\mA_{k_1\ldots k_{M-2}}$.

\subsubsection{Recursive Integration Procedure} 
\label{subsec:algo}

The recursive integration procedure defined by updating through Sections \ref{subsubsec:GMM} and \ref{subsubsec:weighted} is summarized in the supplement. At resolution $m=0$, we obtain $\bpsi_{i,0}(\widehat{\btheta}_{k_1}) = \{\widetilde{\bpsi}_{i,k_1}(\widehat{\btheta}_{k_1})\}_{k_1=1}^{K_1} \in \mathbb{R}^{pK_1}$, $\bPsi_0(\widehat{\btheta}_{k_1}) = \{ \widetilde{\bPsi}_{k_1}(\widehat{\btheta}_{k_1}) \}_{k_1=1}^{K_1} \in \mathbb{R}^{pK_1}$ and a final integrated estimator
\begin{align}
\widehat{\btheta}_r&=\bJ^{-1}_0(\widehat{\btheta}_{k_1}) \bS_0(\widehat{\btheta}_{k_1}) \bV^{-1}_0 (\widehat{\btheta}_{k_1}) 
\Bigl[ \bigl\{ \bS^\top_0(\widehat{\btheta}_{k_1})\bigr\}_j \widehat{\btheta}_j \Bigr]_{j=1}^{K_1},
\label{e:recursive-estimator}
\end{align}
where $\bV_0(\widehat{\btheta}_{k_1})=\sum_{i=1}^N \{ \bpsi_{i,0}(\widehat{\btheta}_{k_1})\}^{\otimes 2}$, $\bS_0(\widehat{\btheta}_{k_1})=- \{ \nabla_{\btheta} \bPsi_0(\btheta) |_{\btheta = \widehat{\btheta}_{k_1}} \}$ and $\bJ_0(\widehat{\btheta}_{k_1})=\bS_0(\widehat{\btheta}_{k_1}) \allowbreak \bV^{-1}_0(\widehat{\btheta}_{k_1}) \bS^\top_0(\widehat{\btheta}_{k_1}) \in \mathbb{R}^{p\times p}$
can be computed following the recursive procedure. 

We give a toy example of the procedure. Consider a $10\times 10$ grid of locations $\mA_0=\{(s_i,s_j)\}_{i,j=1}^{10}$ partitioned into $K_1=2$ regions $\{\mA_1, \mA_2\}$ that are partitioned into $K_2=2$ regions $\{\mA_{11}, \mA_{12}, \mA_{21}, \mA_{22}\}$, illustrated in Figure \ref{toy-example}. We compute $\widehat{\btheta}_r$ as follows:
\begin{enumerate}[leftmargin=*]
\item For $m=2$, compute the kernel score functions $\bpsi_{i,k_1 k_2}(\btheta)$, $i=1, \ldots, N$, to obtain MLEs $\widehat{\btheta}_{k_1k_2}$, $k_1=1,2$, $k_2=1,2$. 
\item For $m=1$, define $\bpsi_{i,k_1}(\widehat{\btheta}_{k_1k_2})=\{ \bpsi^\top_{i,k_11}(\widehat{\btheta}_{k_11}), \bpsi^\top_{i,k_12}(\widehat{\btheta}_{k_12})\}^\top$, $i=1, \ldots, N$, so that we can compute $\bV_{k_1}(\widehat{\btheta}_{k_1k_2})=\sum_{i=1}^N \{\bpsi_{i,k_1}(\widehat{\btheta}_{k_1k_2})\}^{\otimes 2}$. This allows us to estimate the sensitivity with $\bS_{k_1}(\widehat{\btheta}_{k_1k_2})$ and obtain $\widehat{\btheta}_{k_1}$ in \eqref{e:meta-m-1} for $k_1=1,2$.
\item For $m=0$, recursively compute $\bV_0(\widehat{\btheta}_{k_1})=\sum_{i=1}^N [\{ \widetilde{\bpsi}_{i,k_1}(\widehat{\btheta}_{k_1}) \}_{k_1=1}^2]^{\otimes 2}$, $\bS_0(\widehat{\btheta}_{k_1})$, $\bJ_0(\widehat{\btheta}_{k_1})$:
\begin{enumerate}[leftmargin=*]
\item Compute the kernel score functions $\bpsi_{i,k_1k_2}(\widehat{\btheta}_{k_1})$, $i=1, \ldots, N$, $k_1=1,2$, $k_2=1,2$. 
\item Define $\bpsi_{i,k_1}(\widehat{\btheta}_{k_1})=\{ \bpsi^\top_{i,k_11}(\widehat{\btheta}_{k_1}), \allowbreak \bpsi^\top_{i,k_12}(\widehat{\btheta}_{k_1})\}^\top$ so that we can compute $\bV_{k_1}(\widehat{\btheta}_{k_1})=\sum_{i=1}^N \{ \bpsi_{i,k_1}(\widehat{\btheta}_{k_1}) \}^{\otimes 2}$, allowing us to estimate the sensitivity with $\bS_{k_1}(\widehat{\btheta}_{k_1})$, $k_1=1,2$. 
\item Compute $\widetilde{\bpsi}_{i,k_1}(\widehat{\btheta}_{k_1})= \bS_{k_1}(\widehat{\btheta}_{k_1}) \bV^{-1}_{k_1}(\widehat{\btheta}_{k_1}) \allowbreak \bpsi_{i,k_1}(\widehat{\btheta}_{k_1})$, $k_1=1,2$.
\end{enumerate}
Then compute $\bV_0(\widehat{\btheta}_{k_1})$, $\bS_0(\widehat{\btheta}_{k_1})$ and $\bJ_0(\widehat{\btheta}_{k_1})$ to obtain $\widehat{\btheta}_r$ in \eqref{e:recursive-estimator}.
\end{enumerate}

One evaluation at resolution $M$ has computation and memory complexities of $O\{ N(pS_{k_1\ldots k_M})^3 \}$ and $O\{ N(pS_{k_1\ldots k_M})^2 \}$ respectively. This evaluation is repeated $M$ times across the recursive integration procedure. The recursive loop inverts each $(pK_m)\times (pK_m)$ covariance matrix $O(M)$ times, adding computation and memory complexities $O\{ M(pK_m)^3 \}$ and $O\{ M(pK_m)^2 \}$ respectively. At each resolution $m$, inversions and the computation of the $K_m$ estimators $\widehat{\btheta}_{k_1\ldots k_m}$ can be done in parallel across $K_m$ computing nodes to further reduce computational costs. This yields computation and memory complexities, respectively, of 
\scalebox{0.97}{\parbox{\linewidth}{%
\begin{align*}
O\Bigl[\sum \limits_{m=1}^M \bigl\{ N(pS_{k_1\ldots k_M})^3+\max_{k_m=1, \ldots, K_m} NM(pk_m)^3 \bigr\} \Bigr]&=O \Bigl\{ NM(pS_{k_1\ldots k_M})^3+ NM^2(pk_{\max})^3\Bigr\}\\
O\Bigl[\sum \limits_{m=1}^M \bigl\{ N(pS_{k_1\ldots k_M})^2+\max_{k_m=1, \ldots, K_m} NM(pk_m)^2 \bigr\} \Bigr]&=O \bigl\{ NM(pS_{k_1\ldots k_M})^2+ NM^2(pk_{\max})^2\bigr\}.
\end{align*} 
}}\\
Finally, the computation of $\widehat{\btheta}_{k_1\ldots k_m}$ requires no iterative minimization of an objective function, substantially reducing computational costs. The procedure can be fully run on a distributed system, meaning that at no point do the data need to be loaded on a central server or the full $S\times S$ covariance matrix stored.

\subsection{Multi-Resolution Estimating Function Theory}
\label{subsec:asymptotics}

Let $\Theta$ the parameter space of $\btheta$, and denote by $\btheta_0$ the true value of $\btheta$, defined formally by assumptions in the supplement. In this section, we study the asymptotic properties of $\widehat{\btheta}_r$ by formalizing a multi-resolution estimating function theory. To do this, define population versions of the estimating functions, their variability and their sensitivity: for $k_m=1, \ldots, K_m$, $\bphi_{i,k_1\ldots k_m}(\btheta)=\bpsi_{i,k_1\ldots k_m}(\btheta)$ for $m=M,M-1$, and
\begin{align*}
\widetilde{\bphi}_{i,k_1\ldots k_m}(\btheta)&= \bs_{k_1\ldots k_m}(\btheta) \bv^{-1}_{k_1\ldots k_m}(\btheta) \bphi_{i,k_1\ldots k_m}(\btheta), \quad m=M-1, \ldots, 1\\
\bphi_{i,k_1\ldots k_m}(\btheta) &= \{\widetilde{\bphi}_{i,k_1\ldots k_{m+1}}(\btheta) \}_{k_{m+1}=1}^{K_{m+1}}, \quad m=M-2, \ldots, 1,
\end{align*}
where, for $m=1, \ldots, M$, $\bv_{k_1\ldots k_m}(\btheta) = \V_{\btheta_0} \left\{ \bphi_{i,k_1\ldots k_m}(\btheta) \right\}$,
$\bs^\top_{k_1\ldots k_m}(\btheta) = - \E_{\btheta_0} \left\{ \nabla_{\btheta} \bphi_{i,k_1\ldots, k_m}(\btheta) \right\}$ 
and $\bj_{k_1\ldots k_m}(\btheta)=\bs_{k_1\ldots k_m}(\btheta)\allowbreak \bv^{-1}_{k_1\ldots k_m}(\btheta) \bs^\top_{k_1\ldots k_m}(\btheta)$ are the variability, sensitivity and Godambe information \citep{Godambe-1991} matrices, respectively, in $\mA_{k_1\ldots k_m}$. Let $\bphi_{i,0}(\btheta)=\{ \widetilde{\bphi}_{i,k_1}(\btheta) \}_{k_1=1}^{K_1}$, and define $\bv_0(\btheta)=\V_{\btheta_0} \left\{ \bphi_{i,0}(\btheta) \right\}$, $\bs^\top_0(\btheta)=- \E_{\btheta_0} \left\{ \nabla_{\btheta} \bphi_{i,0}(\btheta) \right\} $ and $\bj_0(\btheta)=\bs_0(\btheta) \bv^{-1}_0(\btheta) \bs^\top_0(\btheta)$. We assume throughout that $\bv_{k_1\ldots k_m}(\btheta_0)$ and $\bv_0(\btheta_0)$ are positive definite. 

Under appropriate assumptions on the score functions $\bPsi_{k_1\ldots k_M}(\btheta)$, namely unbiasedness, uniqueness of the root and additivity, the $K_M$ estimators $\widehat{\btheta}_{k_1\ldots k_M}$ from the $M$th resolution are consistent for $\btheta_0$ and semi-parametrically efficient within each $\mA_{k_1\ldots k_M}$, $k_M=1, \ldots, K_M$. Moreover, under appropriate conditions, $\widehat{\btheta}_{k_1\ldots k_{M-1}}$ is consistent and Hansen optimal by \cite{Hector-Song-JASA}. Finally, at each resolution, an optimal estimator is derived from the optimal GMM function \citep{Hansen} and from optimal estimating function theory \citep{Heyde}. This results in a highly efficient integrated estimator $\widehat{\btheta}_r$ in \eqref{e:recursive-estimator} that fully uses the dependence within and between each resolution. This result is shown formally in Theorem \ref{t:cons-norm}.
\begin{theorem}
\label{t:cons-norm}
Under assumptions given in the supplement as $N \rightarrow \infty$, $\widehat{\btheta}_r$ in \eqref{e:recursive-estimator} is consistent and $\sqrt{N} ( \widehat{\btheta}_r - \btheta_0 ) \stackrel{d}{\rightarrow} \mN \{ \bzero, \bj^{-1}_0(\btheta_0) \}$.
\end{theorem}
The proof proceeds by induction after establishing consistency and asymptotic normality of the integrated estimators at resolution $M-1$. It uses a recursive Taylor expansion at each resolution to establish the appropriate convergence rate of the estimating function. Large sample confidence intervals for $\btheta_0$ can be constructed by combining Theorem \ref{t:cons-norm} and the following Corollary, whose proof follows from the proof of Theorem \ref{t:cons-norm} and is omitted.
\begin{corollary}
\label{t:var}
Under assumptions given in the supplement as $N \rightarrow \infty$, $\bJ^{-1}_0(\widehat{\btheta}_r)$ is a consistent estimator of the asymptotic covariance of $\widehat{\btheta}_r$ in \eqref{e:recursive-estimator}.
\end{corollary}

\subsection{Sequential Model Integration Framework}
\label{subsec:implementation}

\begin{figure}[ht]
\begin{tikzpicture}[level distance=1.5cm,
  level 1/.style={sibling distance=6cm},
  level 2/.style={sibling distance=3cm},
  level 3/.style={sibling distance=1cm}]
  \node (A0) {$\mA_0$}
    child {node (A1) {$\mA_1$}
      child {node (A11) {$\mA_{11}$}
      	child {node (A111) {$\mA_{111}$}}
	child {node (A112) {$\mA_{112}$}}
	child {node (A113) {$\mA_{113}$}}
	}
      child {node (A12) {$\mA_{12}$}
      	child {node (A121) {$\mA_{121}$}}
	child {node (A122) {$\mA_{122}$}}
	child {node (A123) {$\mA_{123}$}}
      }
    }
    child {node (A2) {$\mA_2$}
    child {node (A21) {$\mA_{21}$}
    	child {node (A211) {$\mA_{211}$}}
	child {node (A212) {$\mA_{212}$}}
	child {node (A213) {$\mA_{213}$}}}
    child {node (A22) {$\mA_{22}$}
    	child {node (A221) {$\mA_{221}$}}
	child {node (A222) {$\mA_{222}$}}
	child {node (A223) {$\mA_{223}$}}}
    };
    \draw[-latex] (-6.5,-4.5) to[out=180,in=180,looseness=0] (-6.5,-3.5);
    \draw[-latex] (-6.6,-3.5) to[out=135,in=45,looseness=0] (-7.25,-4.5);
    \draw[-latex] (-7.25,-4.5) to[out=180,in=180,looseness=0] (-7.25,-1.8);
    \draw[-latex] (-7.26,-1.8) to[out=135,in=45,looseness=0] (-8,-4.5);
    \draw[-latex] (-8,-4.5) to[out=180,in=180,looseness=0] (-8,-0.25);
    \node (theta11_1r) at (-6.45,-2.8) {\scriptsize $\widehat{\btheta}_{11}$, $\widehat{\btheta}_{12}$,}; 
    \node (theta11_2r) at (-6.5, -3.2) {\scriptsize $\widehat{\btheta}_{21}$, $\widehat{\btheta}_{22}$};
    \node (theta_1r) at (-7.25,-1.5) {\scriptsize $\widehat{\btheta}_1$, $\widehat{\btheta}_2$};
    \node (theta_0r) at (-8,0) {\scriptsize $\widehat{\btheta}_r$};
    
    \draw[-latex](6.5,-4.5) to[out=180,in=180,looseness=0] (6.5,-3.5);
    \draw[-latex](6.5,-2.5) to[out=180,in=180,looseness=0] (6.5,-1.8);
    \draw[-latex](6.5,-1.25) to[out=180,in=180,looseness=0] (6.5,-0.25);
    \node (theta11_1s) at (6.55,-2.8) {\scriptsize $\widehat{\btheta}_{11}$, $\widehat{\btheta}_{12}$,};
    \node (theta11_2s) at (6.5,-3.2) {\scriptsize $\widehat{\btheta}_{21}$, $\widehat{\btheta}_{22}$};
    \node (theta1s) at (6.5,-1.5) {\scriptsize $\widehat{\btheta}_1$, $\widehat{\btheta}_2$};
    \node (theta_0s) at (6.5,0) {\scriptsize $\widehat{\btheta}_s$};
\end{tikzpicture}
\caption{Example partition of $\mA_0$ with $M=3$ resolutions. Schematic propagation steps for computation of $\widehat{\btheta}_{k_1\ldots k_m}$ for recursive (left) and sequential (right) integration schemes.\label{recursive-vs-sequential}}
\end{figure}
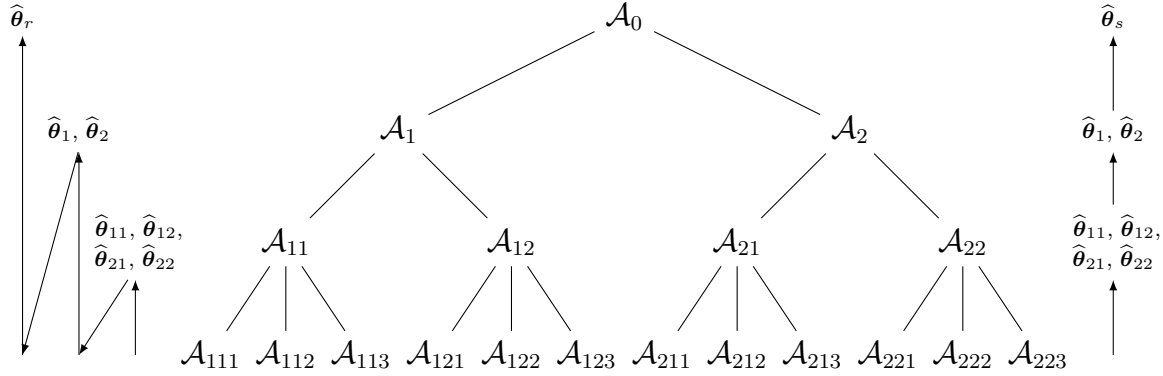

The most time consuming step in the recursive integration procedure requires a recursive update of the weights each time a new estimator of $\btheta$ is computed at resolution $m=M-2, \ldots, 0$, as illustrated on the left of Figure \ref{recursive-vs-sequential}. This is because the weight matrices $\bV_{k_1\ldots k_m}(\widehat{\btheta}_{k_1\ldots k_{m+1}})$ and $\bS_{k_1\ldots k_m}(\widehat{\btheta}_{k_1\ldots k_{m+1}})$ are evaluated at $\widehat{\btheta}_{k_1\ldots k_{m+1}}$. This recursive integration procedure has time complexity which depends on $M^2$, which may be undesirably slow. We propose an alternate sequential integration scheme that evaluates the weight at the estimator from the $M$th resolution, $\widehat{\btheta}_{k_1\ldots k_M}$, a consistent estimator for $\btheta$, illustrated on the right of Figure \ref{recursive-vs-sequential}. The sequential integration procedure replaces the recursive evaluation of the weights with a sequential update by computing $\bV_{k_1\ldots k_m}(\widehat{\btheta}_{k_1\ldots k_M})$ and $\bS_{k_1\ldots k_m}(\widehat{\btheta}_{k_1\ldots k_M})$. In this fashion, no recursive update of the weights is required. We denote by $\widehat{\btheta}_s$ the sequential integrated estimator obtained by evaluating weights using $\bV_{k_1\ldots k_m}(\widehat{\btheta}_{k_1\ldots k_M})$ and $\bS_{k_1\ldots k_m}(\widehat{\btheta}_{k_1\ldots k_M})$. A full algorithm is provided in the supplement.

Arguing as in Section \ref{subsec:algo}, the computation and memory complexities of computing $\widehat{\btheta}_s$ are $O \{ N(pS_{k_1\ldots k_M})^3+ NM(pk_{\max})^3\}$ and $O \{ N(pS_{k_1\ldots k_M})^2+ NM(pk_{\max})^2 \}$ respectively. A consequence of the proof of Theorem \ref{t:cons-norm} is that $\widehat{\btheta}_s$ and $\widehat{\btheta}_r$ are asymptotically equivalent. While $\widehat{\btheta}_s$ may be computationally more advantageous, it requires that the MLEs from resolution $M$ be close to $\btheta_0$, and may not perform as well as $\widehat{\btheta}_r$ in finite samples.

\section{Simulations}
\label{sec:simulations}

We investigate the finite sample performance of the proposed recursive and sequential multi-resolution recursive integrated (MRRI) estimators $\widehat{\btheta}_r$ and $\widehat{\btheta}_s$. Throughout, $\mS$ consists of a square grid of evenly spaced locations. Unless otherwise specified, all simulations are on a Linux cluster using R linked to Intel's MKL libraries with analyses at resolution $M$ performed in parallel across $\widetilde{K}=\prod_{m=1}^M K_m$ CPUs with 1GB of RAM each. Standard errors and confidence intervals are calculated using the results in Theorem \ref{t:cons-norm} and Corollary \ref{t:var}.

In the first set of simulations, we consider a $S=400$-dimensional square spatial domain $\mS= [1,20]^2=\{\bs_j\}_{j=1}^{400}$, $\bs_j \in \mathbb{R}^2$, with $N=10000$. The Gaussian outcomes $\{y_i(\mS)\}_{i=1}^N$ are independently simulated with means $\{\bX^\top_i(\mS) \bbeta\}_{i=1}^N$ and Gaussian spatial covariance function $C_{\alpha}(\bs_j, \bs_{j^\prime}; \bgamma)=\tau^2 \exp\{-\rho^2 (\bs_{j^\prime} - \bs_j)^\top (\bs_{j^\prime} - \bs_j) \}$, with $\bbeta$, $\bgamma=(\tau^2, \rho^2)$ and nugget variance $\sigma^2$ specified below. The covariates $\bX_i(\bs_j)=\bX_i$ consist of an intercept and two non-spatially varying continuous variables independently generated from a Gaussian distribution with mean 0 and variance 4. The true value of the regression coefficient is $\bbeta=(0.3,0.6,0.8)$, the true nugget variance is $\sigma^2=1.6$ and the Gaussian spatial covariance function has true parameters $\tau^2=3$ and $\rho^2=0.5$. To facilitate estimation, we estimate $\btheta=\{\bbeta^\top, \log(\tau^2), \log(\rho^2), \log(\sigma^2)\}$ using $\widehat{\btheta}_r$ and $\widehat{\btheta}_s$. We consider three recursive partitions of $S$ with $M=3$: in Setting I, $K_1=K_2=2, K_3=4$; in Setting II, $K_1=K_3=2$, $K_2=4$; in Setting III, $K_1=4$, $K_2=K_3=2$. Division of the spatial domain is based on nearest neighbours as illustrated in Figure \ref{toy-example}. We also estimate $\btheta$ using two comparative approaches. Specifically, we compare to the partitioning approach (Part.) in \cite{Heaton-etal-2019} that evaluates the sum of the $\widetilde{K}$ log-likelihoods at a grid of values of $\tau^2,\rho^2$, estimates $\tau^2,\rho^2$ using the values that return the highest log-likelihood, then estimates $\bbeta$ and $\sigma^2$ using the least squares estimator and a sample variance respectively; the implementation is modified directly from the code in \cite{Heaton-etal-2019}. Asymptotic standard errors are estimated as the diagonal square root of the inverse variance of the score function. We also compare to the nearest neighbour Gaussian process (NNGP) \citep{Finley-etal-2019} using the 25 nearest neighbours; we implement this ourselves using sparse matrices in Rcpp and parallelize over $\widetilde{K}$ CPUs to make a fair comparison. 
\begin{figure}[h]
\includegraphics[width=\textwidth]{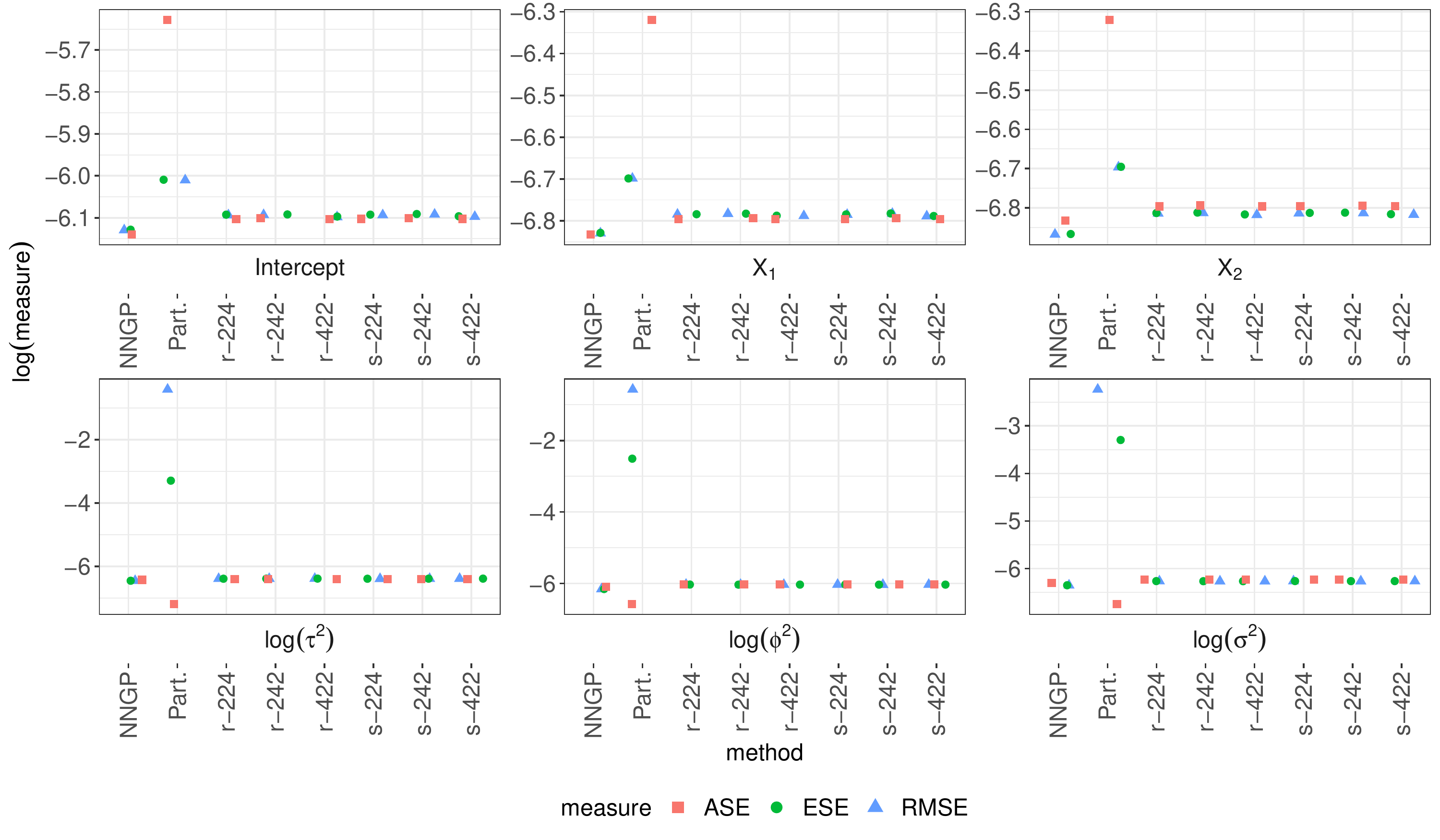}
\caption{RMSE, ESE and ASE for estimates of each parameter using NNGP, Part., $\widehat{\btheta}_r$ and $\widehat{\btheta}_s$ in Settings I, II and III (r$-224$, r$-242$, r$-422$, s$-224$, s$-242$, s$-422$ respectively) for the first set of simulations. \label{f:S400-1}}
\end{figure}
Figure \ref{f:S400-1} plots the root mean squared error (RMSE), empirical standard error (ESE) and asymptotic standard error (ASE) averaged across 500 simulations for each parameter. 

The near equality of RMSE and ESE in Figure \ref{f:S400-1} for our MRRI estimators $\widehat{\btheta}_r$ and $\widehat{\btheta}_s$ illustrates the near unbiasedness of our approach in large samples. Further, the near equality of ESE and ASE justifies the use of the asymptotic variance formula in Theorem \ref{t:cons-norm} and Corollary \ref{t:var} in large samples. Finally, negligible variation is observed across Settings I, II and III, illustrating the robustness of our approach to the chosen recursive or sequential integration scheme. The statistical inference properties of NNGP are similarly favourable, with the NNGP appearing slightly more efficient. Finally, the partitioning approach yields estimators of $\bbeta$ that are nearly unbiased in large samples, whereas the estimators of the covariance parameters show substantial bias. Further, standard errors of the estimators are vastly overestimated for $\bbeta$ and underestimated for the covariance parameters, which is problematic for statistical inference. This is further illustrated in Table \ref{t:S400-CP}, 
\begin{table}[h]
\centering
\caption{95\% confidence interval coverage (CP) in percentage for first set of simulations: $S=400$, $\btheta=\{0.3, 0.6, 0.8, \log(3), \log(0.5), \log(1.6)\}$. \label{t:S400-CP}}
\ra{0.9}
\begin{tabular}{rrrrrrrr}
Method & $K_1,K_2,K_3$ & Intercept & $X_1$ & $X_2$ & $\log(\sigma^2)$ & $\log(\tau^2)$ & $\log(\rho^2)$ \\
\multirow{3}{*}{$\widehat{\btheta}_r$} & 2,2,4 & 94 & 96 & 94 & 96 & 93 & 95 \\ 
& 2,4,2 & 95 & 96 & 94 & 96 & 93 & 95 \\ 
& 4,2,2 & 94 & 95 & 95 & 96 & 93 & 95 \\ 
\multirow{3}{*}{$\widehat{\btheta}_s$} & 2,2,4 & 94 & 96 & 94 & 96 & 93 & 95 \\ 
& 2,4,2 & 95 & 96 & 94 & 96 & 93 & 95 \\ 
& 4,2,2 & 94 & 95 & 95 & 96 & 93 & 95 \\ 
Part. & & 100 & 100 & 100 & 0 & 0 & 0 \\ 
NNGP & & 95 & 95 & 95 & 96 & 96 & 96 \\ 
\end{tabular}
\end{table}
which reports the 95\% confidence interval coverage (CP) for all estimators averaged across 500 simulations. Whereas CP reaches nominal levels across parameters for MRRI and NNGP, regression and covariance parameters are respectively overcovered and undercovered using Part.

Computing times in seconds are reported in Table \ref{t:S400-time}. 
\begin{table}[h]
\centering
\caption{Mean elapsed time (Monte Carlo standard deviation) in seconds of $\widehat{\btheta}_r$, $\widehat{\btheta}_s$, NNGP and Part. across 500 simulation for the first set of simulations: $S=400$, $\btheta=\{0.3, 0.6, 0.8, \log(3), \log(0.5), \log(1.6)\}$. \label{t:S400-time}}
\ra{0.9}
\begin{tabular}{rrrrr}
Method & $K_1,K_2,K_3=2,2,4$ & $K_1,K_2,K_3=2,4,2$ & $K_1,K_2,K_3=4,2,2$ & 16 CPUs\\
$\widehat{\btheta}_r$ & 0.80 (0.092) & 0.66 (0.060) & 0.66 (0.060) &\\
$\widehat{\btheta}_s$ & 0.72 (0.084) & 0.60 (0.064) & 0.60 (0.064) & \\ 
NNGP & & & & 280 (140)\\
Part. & & & & 127 (11)
\end{tabular}
\end{table}
Recursively updating weights with $\widehat{\btheta}_r$ is faster than using the weights computed at the $M$th layer with $\widehat{\btheta}_s$. In comparison, the partitioning and NNGP approaches are $212$ and $467$ times slower, respectively, than our estimator $\widehat{\btheta}_s$ with $K_1=4,K_2=K_3=2$.

In the second set of simulations, we consider a $S=25600$-dimensional square spatial domain $\mS=[1,160]^2=\{\bs_j\}_{j=1}^{25600}$, $\bs_j \in \mathbb{R}^2$, with $N=5000$ in Setting I and $N=2000$ in Setting II. The covariates and outcomes are generated as in the first set of simulations. The true value of $\btheta$ is the same as the first set of simulations, $M=4$, $K_1=K_2=K_3=K_4=4$. Division of the spatial domain is based on nearest neighbours as illustrated in Figure \ref{toy-example}. Analyses are parallelized across 16 CPUs with 2GB of RAM each. Given the performance of the partitioning approach in the first set of simulations, we only attempted to compare to the NNGP using the 100 nearest neighbours parallelized over $16$ CPUs. None of the 500 simulations finished before timing out at 24 hours. Table \ref{t:S25600-1} 
\begin{table}[h]
\centering
\caption{Simulation metrics of $\widehat{\btheta}_r$ and $\widehat{\btheta}_s$ across 500 simulations for the second set of simulations: $S=25600$, $\btheta=\{0.3, 0.6, 0.8, \log(3), \log(0.5), \log(1.6)\}$. \label{t:S25600-1}}
\ra{0.9}
\begin{subtable}[H]{\textwidth}
\caption{Simulation metrics in Setting I.}
\begin{tabular}{llrrrrr}
Estimator & Parameter & RMSE$\times 10^4$ & ESE$\times 10^4$ & ASE$\times 10^4$ & BIAS$\times 10^5$ & CP $(\%)$ \\ 
\multirow{6}{*}{$\widehat{\btheta}_r$} & Intercept & 4.2 & 4.2 & 4.1 & $-2.7$ & 96 \\ 
& $\bX_1$ effect & 2.0 & 2.0 & 2.1 & 0.021 & 96 \\ 
& $\bX_2$ effect & 2.1 & 2.1 & 2.1 & 0.42 & 94 \\ 
& $\log(\tau^2)$ & 2.7 & 2.7 & 2.8 & $-1.7$ & 96 \\ 
& $\log(\rho^2)$ & 4.1 & 4.0 & 3.7 & 8.7 & 93 \\ 
& $\log(\sigma^2)$ & 3.2 & 3.1 & 3.1 &$-3.3$ & 96 \\ 
\multirow{6}{*}{$\widehat{\btheta}_s$} & Intercept & 4.2 & 4.2 & 4.1 & $-2.7$ & 96 \\ 
& $\bX_1$ effect &2.0 & 2.0 & 2.0 & 0.073 & 96 \\ 
& $\bX_2$ effect & 2.1 & 2.1 & 2.0 & 0.42 & 94 \\ 
& $\log(\tau^2)$ & 2.7 & 2.7 & 2.8 & $-2.2$ & 96 \\ 
& $\log(\rho^2)$ & 4.1 & 4.0 & 3.7 & 9.4 & 93 \\ 
& $\log(\sigma^2)$ & 3.2 & 3.1 & 3.1 & $-3.5$ & 95 \\ 
\end{tabular}
\end{subtable}\\
\begin{subtable}[H]{\textwidth}
\caption{Simulation metrics in Setting II.}
\begin{tabular}{llrrrrr}
Estimator & Parameter & RMSE$\times 10^4$ & ESE$\times 10^4$ & ASE$\times 10^4$ & BIAS$\times 10^4$ & CP $(\%)$ \\ 
\multirow{6}{*}{$\widehat{\btheta}_r$} & Intercept & 6.5 & 6.5 & 6.5 & $-0.69$ & 95 \\ 
& $\bX_1$ effect & 3.4 & 3.4 & 3.2 & $-0.073$ & 93 \\ 
& $\bX_2$ effect & 3.4 & 3.4 & 3.2 & 0.044 & 94 \\ 
& $\log(\tau^2)$ & 4.5 & 4.5 & 4.4 & $-0.28$ & 94 \\ 
& $\log(\rho^2)$ & 6.4 & 6.3 & 5.9 & 1.3 & 93 \\ 
& $\log(\sigma^2)$ & 5.1 & 5.1 & 4.8 & $-0.40$ & 92 \\ 
\multirow{6}{*}{$\widehat{\btheta}_s$} & Intercept & 6.5 & 6.5 & 6.4 & $-0.72$ & 94 \\ 
& $\bX_1$ effect & 3.4 & 3.4 & 3.2 & $-0.065$ & 93 \\ 
& $\bX_2$ effect & 3.4 & 3.4 & 3.2 & 0.057 & 94 \\ 
& $\log(\tau^2)$ & 4.5 & 4.5 & 4.4 & $-0.39$ & 94 \\ 
& $\log(\rho^2)$ & 6.4 & 6.3 & 5.9 & 1.4 & 93 \\ 
& $\log(\sigma^2)$ & 5.1 & 5.1 & 4.8 & $-0.43$ & 92 \\ 
\end{tabular}
\end{subtable}
\end{table}
reports the RMSE, ESE, ASE, mean bias (BIAS) and CP of our MRRI estimators averaged across 500 simulations.

Again, simulation metrics in Table \ref{t:S25600-1} support the use of Theorem \ref{t:cons-norm} and Corollary \ref{t:var} in finite samples: the RMSE, ESE and ASE are approximately equal, and the BIAS is negligible. We observe appropriate CP, with a slight undercoverage in Setting II with smaller sample size. Estimators $\widehat{\btheta}_r$ and $\widehat{\btheta}_s$ appear equivalent in these large sample size settings. The desirable statistical performance observed in the first set of simulations remains as we increase the number of resolutions, $M$. Finally, mean elapsed times (Monte Carlo standard deviation) in seconds are $390$ ($170$) and $240$ ($110$) for $\widehat{\btheta}_r$ in Settings I and II respectively, and $380$ ($140$) and $230$ ($89$) for $\widehat{\btheta}_s$ in Settings I and II respectively.

The third set of simulations mimics the data analysis of Section \ref{sec:application} with $S=800$. We consider two ROIs, $\mS_1=[1,20]^2=\{\bs_j\}_{j=1}^{400}$ and $\mS_2=[21,40]^2=\{\bs_j\}_{j=401}^{800}$, $\bs_j \in \mathbb{R}^2$. Defining $\mS=\mS_1 \cup \mS_2$, the Gaussian outcomes $\{y_i(\mS)\}_{i=1}^N$, $N=10000$, are independently simulated with mean $\{\bone \beta\}_{i=1}^N$, $\beta=0$, $\bone \in \mathbb{R}^S$ a vector of one's, and the spatial covariance function in \eqref{e:ABIDE-cov} with $d=2$. The spatial variance is modeled through $\tau(\bs_j, \bs_{j^\prime})=\tau^2$, the spatial correlation is modeled through $\brho(\bs_j) = \brho_1 \mathbbm{1}(\bs_j \in \mS_1) + \brho_2 \mathbbm{1}(\bs_j \ \in \mS_2)$, and $\bgamma=\{\log(\tau^2), \brho^\top_1, \brho^\top_2\}^\top \in \mathbb{R}^{1+2q}$. Here, $\bX_i \in \mathbb{R}^q$ consists of an intercept and two non-spatially varying continuous variables independently generated from a Gaussian distribution with mean $0$ and variance $1$. The true values of the dependence parameters are set to $\sigma^2=1.6$, $\tau^2=3$, $\brho_1=(0.5,0.5,0.5)$ and $\brho_2=(0.6,0.6,0.6)$. We estimate $\btheta=\{\beta, \log(\tau^2), \brho^\top_1, \brho^\top_2, \log(\sigma^2)\}$ using $\widehat{\btheta}_s$ using the recursive partition of $S$ with $M=3$: $K_1=K_2=2$, $K_3=4$. Each set $\mA_{k_1\ldots k_m}$ at resolution $m$ is a union of the nearest neighbours in $\mS_1$ and $\mS_2$ separately, so that each set $\mA_{k_1k_2k_3}$ consists of $25$ locations from $\mS_1$ and $25$ locations from $\mS_2$, $\mA_{k_1k_2}$ consists of $100$ locations from $\mS_1$ and $100$ locations from $\mS_2$, and so on. Analyses are parallelized across 16 CPUs with 2GB of RAM each. Table \ref{t:S800-1} 
\begin{table}[h]
\centering
\caption{Simulation metrics of $\widehat{\btheta}_s$ across 500 simulations for the third set of simulations: $S=800$, $\btheta=\{0,\log(3),0.5,0.5,0.5,0.6,0.6,0.6,\log(1.6)\}$. \label{t:S800-1}}
\ra{0.9}
\begin{tabular}{lrrrrr}
Parameter & RMSE$\times 10^3$ & ESE$\times 10^3$ & ASE$\times 10^3$ & BIAS$\times 10^4$ & CP $(\%)$ \\ 
$\beta$ & $1.4$ & $1.4$ & $1.4$ & $-0.78$ & $94$ \\ 
$\log(\tau^2)$ & $1.1$ & $1.1$ & $1.1$ & $0.44$ & $96$ \\ 
$\rho_{11}$ & $1.9$ & $1.9$ & $1.9$ & $-0.57$ & $94$ \\ 
$\rho_{12}$ & $1.9$ & $1.9$ & $1.8$ & $-0.88$ & $94$ \\ 
$\rho_{13}$ & $1.9$ & $1.8$ & $1.8$ & $-1.80$ & $93$ \\ 
$\rho_{21}$ & $1.9$ & $1.9$ & $1.9$ & $-0.35$ & $94$ \\ 
$\rho_{22}$ & $1.9$ & $1.9$ & $1.9$ & $-1.20$ & $95$ \\ 
$\rho_{23}$ & $1.9$ & $1.9$ & $1.9$ & $-0.49$ & $94$ \\ 
$\log(\sigma^2)$ & $1.2$ & $1.2$ & $1.2$ & $-0.51$ & $96$ \\ 
\end{tabular}
\end{table}
reports the RMSE, ESE, ASE, BIAS and CPU of our MRRI estimator $\widehat{\btheta}_s$ averaged across $500$ simulations, where $\brho_j=(\rho_{j1},\rho_{j2},\rho_{j3})$, $j=1,2$.

Parallelizing over $\widetilde{K}=16$ CPUs, mean elapsed time (Monte Carlo standard deviation) is $29$ minutes ($22$ minutes). Simulation metrics in Table \ref{t:S800-1} are consistent with the results from the previous simulations: the RMSE, ESE and ASE are approximately equal, the BIAS is negligible, and the CP reaches the nominal 95\% level. Further, to illustrate the high statistical power of our approach, we perform a test of the hypotheses $H_0: \rho_{12} = \rho_{22}$ versus $H_A: \rho_{12} \neq \rho_{22}$ using the test statistic $Z=(\widehat{\rho}_{12} - \widehat{\rho}_{22} -\rho_0)\{\V(\widehat{\rho}_{12}) + \V(\widehat{\rho}_{22}) - 2\mbox{Cov}(\widehat{\rho}_{12}, \widehat{\rho}_{22})\}^{-1/2}$,
which follows an approximate standard normal distribution under $H_0$ when $\rho_0=0$. In the context of the analysis of Section \ref{sec:application}, this test evaluates whether ASD is associated with different spatial correlation in the left and right precentral gyri. Across the $500$ simulations, the test rejects the null 100\% of the time at level $0.05$, illustrating the high statistical power of our approach. Using $\rho_0=\rho_{12}-\rho_{22}$, the type-I error rate is $3.6\%$ across the 500 simulations.

\section{Estimation of Brain Functional Connectivity}
\label{sec:application}

We return to the motivating neuroimaging application described in Section \ref{sec:introduction}. Out of 1112 ABIDE participants, $774$ passed quality control: 379 with ASD, 647 males (335 with ASD) and 127 females (44 with ASD), with mean age 15 years (standard deviation 6 years). The left and right precentral gyri form the primary motor cortex and are responsible for executing voluntary movements \citep{Bookheimer}. They are two of the largest ROIs in the Harvard-Oxford atlas \citep{FSL} with 1786 and 1888 voxels respectively. Many individuals with ASD have motor deficits \citep{Jansiewicz-etal}. Atypical connectivity within the left and right precentral gyri may indicate that these motor deficits are related to how these two brain regions coordinate movement. The pre-processing pipeline of rfMRI data has already been described by \cite{ABIDE}.  Participant-specific data have been registered into a common template space such that voxel locations are comparable between participants in the study. 

Define $\mS_1=\{\bs_j\}_{j=1}^{1888}$ and $\mS_2=\{\bs_j\}_{j=1889}^{3674}$ the set of voxels in the right and left precentral gyri, respectively, $d=3$ the dimension of $\bs_j$, and $\mS=\mS_1 \cup \mS_2$. Voxels in the atlas outside the brain are assigned missing. Following the thinning described in Section \ref{subsec:setup}, we obtain independent replicates $y_i(\bs_j)$ observed at $S=3674$ voxel locations, $i=1, \ldots, 75888$. We refer to this dataset as the ``primary'' dataset; as an evaluation of the robustness of our estimation approach, we compare estimates from this primary dataset to estimates from a secondary dataset, consisting of the excluded time points, with identical dimensions $N$ and $S$. There is \emph{a priori} no reason to believe the data distribution differs across primary and secondary datasets, and so comparing results across both datasets will allow us to quantify robustness of our analysis, a notoriously difficult task in analyses of rfMRI data \citep{Uddin-etal-2017}.

Let $\bX_i$ be corresponding observations of $q=5$ covariates for outcome $i$: an intercept, ASD status (1 for ASD, 0 for neurotypical), age (centered and standardized), sex (0 for male, 1 for female) and the age (centered and standardized) by ASD status interaction. We model $\mu(\bs_j; \bX_i, \bbeta)=\beta$, where $\mu(\bs_j; \bX_i, \bbeta)$ is the mean rfMRI time series at voxel $\bs_j$. The covariance $C_{\alpha}(\bs_{j^\prime}, \bs_j; \bX_i, \bgamma)$ is that given in equation \eqref{e:ABIDE-cov} of Section \ref{subsec:setup} with $d=3$ and $\tau(\bs_j, \bs_{j^\prime})= \{ (\tau^2_1)^{\mathbbm{1}(\bs_j \in \mS_1) + \mathbbm{1}(\bs_{j^\prime} \in \mS_1)} (\tau^2_2)^{\mathbbm{1}(\bs_j \in \mS_2) + \mathbbm{1}(\bs_{j^\prime} \in \mS_2)} \}^{1/2}$. As in Section \ref{sec:simulations}, the correlation is modeled through $\brho(\bs_j) = \brho_1 \mathbbm{1}(\bs_j \in \mS_1) + \brho_2 \mathbbm{1}(\bs_j \ \in \mS_2)$. Thus, $\bgamma=\{\log(\tau^2_1), \log(\tau^2_2), \brho_1^\top, \brho_2^\top\} \in \mathbb{R}^{2+2q}$ and $\btheta=\{\beta, \bgamma, \log(\sigma^2)\}$, with $\bgamma$ the parameter of primary interest describing the effect of covariates on functional connectivity within the left and precentral gyri. Correlation between the two ROIs is leveraged through the multivariate approach for increased precision. We interpret the effect of ASD on the correlation structure in detail in the supplement. Within each ROI, the ASD effect only influences the rate of the decay of the spatial correlation in the exponential term. An illustration of the correlation between ROIs for various values of $\rho_{j2}, \rho_{j^\prime 2}$ is provided in the supplement.
 
The size of each (primary and secondary) outcome dataset is 15GB. To overcome the computational burden of a whole dataset analysis, we estimate $\btheta$ using the sequential estimator $\widehat{\btheta}_s$. To partition the three-dimensional spatial domain, we recursively partition $\mS_1$ and $\mS_2$ separately into $K_1=2$, $K_2=K_3=4$ disjoint sets based on nearest neighbours, $M=3$. The sets $\mA_{k_1\ldots k_m}$ consist of the union of the disjoint partition sets of $\mS_1$ and $\mS_2$ at each resolution $m \in \{1, \ldots, M\}$. A plot of the partitioning is provided in the supplement.

The analysis of the primary and secondary datasets takes $5.2$ hours each when parallelized across $32$ CPUs. The estimated effect and standard deviation (s.d.) of the intercept, ASD status, age, sex and age by ASD status interaction are reported in Table \ref{t:ABIDE-estimates}. In the primary dataset, the estimates (s.d.) of $\beta$, $\log(\sigma^2)$, $\log(\tau^2_1)$ and $\log(\tau^2_2)$ are, respectively, $-0.00538$ ($2.03\times10^{-4}$), $-4.05$ ($3.90\times 10^{-4}$), $-0.108$ ($2.53\times 10^{-4}$), $-0.0699$ ($2.48\times 10^{-4}$). In the secondary dataset, the estimates (s.d.) of $\beta$, $\log(\sigma^2)$, $\log(\tau^2_1)$ and $\log(\tau^2_2)$ are, respectively, $-0.00606$ ($2.03\times10^{-4}$), $-4.05$ ($3.88\times10^{-4}$), $-0.111$ ($2.51\times10^{-4}$), $-0.0711$ ($2.46\times10^{-4}$). As expected, estimates of $\beta$ are close to $0$ and estimates of $\sigma^2+\tau^2_j$ are close to 1 due to the centering and standardizing. We measure the agreement between standardized estimates of $\btheta$ in the primary and secondary datasets using cosine similarity, the cosine of the angle between the two standardized vectors $\widehat{\theta}_{q}/\{\V(\widehat{\theta}_{q}) \}^{1/2}$, $q=1, \ldots, 14$. The cosine similarity is $0.999995$, indicating a high degree of agreement between the standardized vectors.

\begin{table}[h]
\centering
\caption{Estimated covariate effects and s.d. in the primary and secondary datasets. \label{t:ABIDE-estimates}}
\ra{0.9}
\begin{tabular}{rrrrr}
& \multicolumn{2}{c}{$\brho_1$} &  \multicolumn{2}{c}{$\brho_2$}\\
Covariate & estimate & s.d.$\times 10^4$ & estimate & s.d.$\times 10^4$ \\ 
& \multicolumn{4}{c}{primary dataset} \\
Intercept & $0.569$ & $1.44$ & $0.561$ & $1.42$\\
ASD status & $-0.00628$ & $1.32$ & $-0.0046$ & $1.27$\\
age & $-0.0221$ & $1.07$ & $-0.0119$ & $0.932$\\
sex &  $0.0362$ & $1.79$ & $0.0656$ & $1.72$\\
age by ASD status interaction & $-0.00327$ & $1.38$ & $-0.00397$ & $1.25$\\
& \multicolumn{4}{c}{secondary dataset} \\
Intercept & $0.568$ & $1.43$ & $0.562$ & $1.41$\\
ASD status & $-0.0055$ & $1.32$ & $-0.00509$ & $1.27$\\
age & $-0.0211$ & $1.07$ & $-0.0118$ & $0.931$\\
sex & $0.0342$ & $1.78$ & $0.0642$ & $1.72$\\
age by ASD status interaction & $-0.00492$ & $1.38$ & $-0.00586$ & $1.25$
\end{tabular}
\end{table}

From a practical perspective, estimates and their standard errors are virtually identical across primary and secondary datasets. Two-sample $Z$-tests, however, mostly reject the null that elements of $\btheta$ are equal in both datasets at the typical $0.05$ level: this is a feature of the sample size and the high power of the test, rather than of true underlying differences between the two datasets, and illustrates well the challenges of robustness in analyses of rfMRI data. We calibrate the $\alpha$-level of hypothesis testing procedures in our analysis by borrowing ideas from knock-offs \citep{Barber-Candes}. We estimate a data-dependent type-I error rate threshold as the $5$\% quantile of the observed $p$-values of the two-sample tests of equality between parameters in primary and secondary datasets. The Gaussian critical value corresponding to this $5$\% quantile is $z_{\mbox{\scriptsize{crit}}}=8.60$. 

Armed with this robust critical value, we evaluate whether the ASD main and interaction effects are significantly different across the two brain regions. We perform a test of the hypotheses $H^m_0: \rho_{12}=\rho_{22}$ versus $H^m_A: \rho_{12} \neq \rho_{22}$ and $H^i_0: \rho_{15}=\rho_{25}$ versus $H^i_A: \rho_{15}\neq \rho_{25}$ using the test statistic $Z$ in Section \ref{sec:simulations}, where $\rho_{j2}$ and $\rho_{j5}$ are the ASD main and interaction effects, respectively, in ROI $\mS_j$, $j=1,2$. The test statistic for $H^m_0$ versus $H^m_A$ takes a value of $13.3$ and $3.23$ in the primary and secondary datasets respectively; since we reject $H^m_0$ in the primary but not the secondary dataset, we conclude that the main ASD effect is not significantly different in the correlation structure of both brain regions. The test statistic for $H^i_0$ versus $H^i_A$ takes a value of $5.70$ and $7.62$ in the primary and secondary datasets respectively, and we conclude that the ASD by age interaction effect is not significantly different in the correlation structure of both brain regions. An analysis that excludes the age by ASD status interaction is included in the supplement and agrees with this analysis.

Summaries of distances $d_{jj^\prime}=(\bs_{j^\prime} - \bs_j)^\top (\bs_{j^\prime} - \bs_j)$ are provided in the supplement. For a male participant of mean age ($15.13$ years), the estimated correlation structures between the right and left precentral gyri, within the right precentral gyrus, and within the left precentral gyrus, for a participant with and without ASD are, respectively, $0.915 \exp ( - 1.83 d_{jj^\prime} )$ and $0.915 \exp ( - 1.86 d_{jj^\prime} )$, $0.898 \exp ( - 1.85 d_{jj^\prime} )$ and $0.898 \exp ( - 1.89 d_{jj^\prime} )$, and $0.933 \exp ( - 1.82 d_{jj^\prime} )$ and $0.933 \exp ( - 1.84 d_{jj^\prime} )$. ASD manifests as hyper-connectivity within and between the right and left precentral gyri. These findings concur with those of \cite{Nebel-etal}. These results also concur with a less powered analysis that averages each participant's rfMRI times series at each voxel, then regresses the participants' Pearson correlation between the right and left precentral gyri onto an intercept, ASD status, age, sex and the age by ASD status interaction. Estimates (s.d.) of covariate effects from this analysis are $0.726$ ($0.00759$), $-0.0457$ ($0.0101$), $0.0159$ ($0.00745$), $-0.0111$ ($0.0137$), $-0.00293$ ($0.0101$). Only the intercept, ASD status and age effects are significant at level $0.05$, highlighting the superior power of our spatial approach.

\section{Discussion}
\label{sec:discussion}

The proposed recursive and sequential integration estimators depend on the choice of the recursive partition of $\mS$. We have suggested that this partitioning be performed such that $S_{k_1\ldots k_m}$ and $K_m$, $m=1, \ldots, M$ are relatively small compared to $S_0$ and shown through simulations that this leads to desirable statistical and computational performance. Nonetheless, the GMM is known to underestimate the variance of estimators when $K_m$ is moderately large relative to $N$; see \cite{Hansen-Heaton-Yaron} and others in the special issue. Thus, special care should be taken to ensure $K_m$ is relatively low-dimensional.

In this paper, we have allowed $\btheta$ to vary spatially under the constraint that it can be consistently estimated using subsets of the spatial domain $\mS$. Other settings may consider a setting in which each subset is modeled through its own, subset-specific parameter. Future research should focus on the development of recursive and sequential integration rules for this setting with partially heterogeneous parameters $\btheta$ following, for example, the work of \cite{Hector-Reich} in spatially varying coefficient models. \emph{A priori}, these extensions should follow from zero-padding the weight matrices in equations \eqref{e:meta-m-1} and \eqref{e:recursive-estimator} and implementing an accounting system to keep track of the heterogeneous and homogeneous model parameters, although a thorough investigation needs to be performed to validate this extension.

While our approach was motivated by a comparison of functional connectivity between participants with and without ASD, the proposed methods are generally applicable to Gaussian process modeling of high-dimensional images. Our ABIDE analysis has primarily focused on the association between ASD and correlation within the left and right precentral gyri, where correlation between the two regions was modeled as a function of the within-ROI correlations. This analysis is suitable when within-ROI correlation is of primary interest. Extensions that model the between-ROI correlation through a more flexible covariance structure are of interest but beyond the scope of the present work.

\section*{Acknowledgements}

The authors are grateful to the participants of the ABIDE study, and the ABIDE study organizers and members who aggregated, preprocessed and shared the ABIDE data.

\appendix

\bibliographystyle{apalike}
\bibliography{RecursiveIntegration-bib-20210218}

\end{document}